\documentclass[pdflatex,sn-nature]{sn-jnl}

\usepackage{graphicx}%
\usepackage{multirow}%
\usepackage{amsmath,amssymb,amsfonts}%
\usepackage{amsthm}%
\usepackage{mathrsfs}%
\usepackage[title]{appendix}%
\usepackage{xcolor}%
\usepackage{textcomp}%
\usepackage{manyfoot}%
\usepackage{booktabs}%
\usepackage{algorithm}%
\usepackage{algorithmicx}%
\usepackage{algpseudocode}%
\usepackage{listings}%
\usepackage{svg}
\usepackage[T1]{fontenc}
\usepackage[utf8]{inputenc}
\usepackage{amssymb}
\usepackage{blindtext}
\usepackage{float}
\usepackage{bm}

\theoremstyle{thmstyleone}%
\theoremstyle{thmstyletwo}%

\theoremstyle{thmstylethree}%

\begin{document}

\title[Article Title]{Fully Coupled Multiphysics Modelling of Fracture Behaviour in Silicon Particles during Lithiation–Delithiation Using the Phase-Field Method}


\author [1]{\fnm{Jie} \sur{Yang}}\email{exx845@qmul.ac.uk}

\author*[1]{\fnm{Wei} \sur{Tan}}\email{wei.tan@qmul.ac.uk}

\affil [1]{\orgdiv{School of Engineering and Materials Science}, \orgname{Queen Mary university of London}, \orgaddress{\street{Mile End Road}, \city{London}, \postcode{E14NS}, \country{UK}}}


\abstract{In this study, a multiphysics model fully coupling mass transport, deformation, phase field, and fatigue damage was developed to investigate the cracking and fracturing behaviours of Si particles during the single lithiation-delithiation cycle and fatigue damage during multiple cycles. The effects of particle diameter, charge rate, and pre-existing notches on the failure behaviour of Si particles were systematically analysed. The results showed that the increase in charge rate, particle diameter, and pre-existing notch length leads to larger cracking rates and faster fracturing of the particle. Then, a validated contour map of Si particle’s fracture behaviours was developed. Additionally, the influence of pre-existing notch length and charge rate on fatigue damage was examined, and it was found that longer pre-existing notch length and larger charge rate can shorten the particle's cyclic life. Finally, to alleviate the particle fracture, nanopores were introduced in the particle, and the influence of porosity on the fracture behaviours was investigated. The results showed that nanopores can reduce expansion, dissipate global tensile stresses and elongate the crack propagation path. The developed computational framework established a predictive relationship between stress-diffusion coupling theory and particle-level degradation, guiding future design and manufacturing of failure-resistant Si-based anodes for lithium-ion batteries.}




\maketitle

\section{Introduction}

\label{sec1} 

The rapid development of electric vehicles has led to an increasing
demand for longer driving ranges, which, combined with the requirement
for lightweight designs, necessitates the development of electrodes
with higher energy densities  \citep{he2021challenges,wang2022toward}.
While the energy density potential of cathode materials, such as lithium
nickel manganese cobalt oxides (NMC), is approaching its theoretical
limit, anode materials currently exhibit greater potential for advancements
\citep{cheng2021recent}. Silicon-based materials have emerged as
a promising alternative due to their exceptionally high theoretical
lithium-ion capacity (3579 mAh/g for Li\textsubscript{1}\textsubscript{5}Si\textsubscript{4}),
which is nearly ten times of that of graphite (372 mAh/g) \citep{sun2022recent,hossain2024silicon}.
However, silicon particles undergo approximately 300\% expansion during
lithiation \citep{lee2020stress,ma2020strategic}, and repeated lithiation/delithiation
cycles result in excessive cyclical volume changes, leading to particle
cracking, fracturing, and ultimately electrode failure \citep{meng2020internal,ge2021recent,kim2023issues}.
To investigate the fracturing behaviours of silicon particles, Liu,
et al. \citep{liu2012size} and Tokranov et al. \citep{tokranov2014situ}
employed in-situ transmission electron microscopy (TEM) and atomic
force microscopy (AFM) techniques, respectively, to study the influence
of particle diameters. Their findings revealed that particles smaller
than 150 nm did not exhibit cracking.

Despite these advances, the failure mechanism of silicon particles
remains unclear due to the limitations of experimental methods in
capturing fundamental characteristics such as stress and concentration
distributions. To address these challenges, simulation methods have
been employed. For instance,
Huang et al. \citep{huang2013stress} used an elastic-perfectly plastic
model to describe lithiation-induced deformation and concluded that
particle size influenced fracturing through strain energy release. Zhao and his colleagues studied the lithiation of crystalline silicon via simulations and experiments, illustrating the particle deformation and stress evolution during the initial lithiation \citep{zhao2011concurrent,zhao2011lithium,pharr2012kinetics}. And recently, Schoof et al. studied the stress evolution and distribution of amorphous Si during single cycle and multiple cycles \citep{schoof2025modeling}. Although these studies only provided stress distribution insights without describing particle damage, they provide solid foundation for the studies of cracking and fracturing behaviours of Si particle during (de)lithiation.

In recent years, the phase-field method has gained prominence in solid
damage modelling. Based on Griffith’s energy-based fracture theory,
this method treats evolving interfaces as diffuse rather than sharp
and can effectively capture crack propagation \citep{zhuang2022phase,li2023review}.
Additionally, its compatibility allows seamless coupling with other
physics, like corrosion \citep{kovacevic2023phase}, microstructure
evolution \citep{tourret2022phase}, mechanical fracture \citep{simoes2021phase},
thermal spalling \citep{cheng2022coupled}, et al., making it ideal
for multiphysics modelling. Xu et al. \citep{liu2020cracks} developed
a multiphysics model using the phase-field method to describe crack
initiation and propagation in silicon nanoparticles. They demonstrated
that particle size and lithiation rate were dominant factors in cracking
and proposed a quantitative relationship between these variables for
crack initiation. Similarly, Choi et al. \citep{choi2023phase} combined
a phase-field damage model, a Li-ion diffusion model, and a mechanical
model to characterise crack initiation and propagation in silicon
nanoparticles, identifying mismatched stress/strain due to concentration
gradients as the primary cause of damage.

Nevertheless, these models have primarily focused on electrode particles
under single charge-discharge steps, whereas in practice, the particles
experience cyclic loading and fatigue damage. Chew et al. \citep{chew2014cracking}
used a finite element model incorporating pre-existing microcracks
to examine the combined effects of finite strains, plastic flow, and
pressure gradients on lithium diffusion during repeated lithiation/delithiation
cycles. Ai et al. \citep{ai2022coupled} employed a multiphysics phase-field
model to capture fatigue crack propagation in NMC cathode particles,
exploring the effects of C-rate, particle size and crack geometry
on damage. Zhang et al. \citep{zhang2023cycling} constructed a mechanochemically
coupled model to analyse cycling-induced stress evolution, mechanical
degradation and capacity loss, proposing an optimal charging protocol
to mitigate damage.

However, research gaps remain in existing studies. Experimental results
have shown that silicon particles undergo a crystalline-to-amorphous
phase transformation during lithiation, altering their mechanical
properties \citep{ogata2014revealing,shenoy2010elastic,choi2015simple,shi2016failure,khosrownejad2017crack}
- a factor often neglected in prior models. Furthermore, previous
studies have demonstrated that pre-existing micro-flaws significantly
influence fracture behaviour \citep{chew2014cracking}. However, despite
their inevitability in practical applications, these effects have
not been extensively investigated. Lastly, a porous structure is good
 for accommodating expansion, thus likely to reduce particle fracturing
\citep{zhang2021challenges}. While porous silicon particles have
been proposed \citep{ge2013scalable,wang2015monodisperse,jia2018novel,yan2022scalable},
their influence mechanism and the effects of porosity on fracture
behaviour remain largely unexplored. To bridge these gaps, in this
work, a multiphysics model coupling mass transport, solid mechanics,
and phase-field damage was developed and validated against experimental
results. Fatigue damage was incorporated into the phase-field damage
model to simulate cyclic lithiation/delithiation processes. The modelling
details are presented in Section 2. Section 3 discusses the mechanism
of silicon particle fracturing, followed by studies on particle diameter,
charge rate effects, pre-existing notches, fatigue damage under low
charge rates, and the influence of nanopores.


\section{Multiphysics model}

The intercalation and extraction of lithium (Li) ions into and out
of silicon (Si) particles are described as diffusion processes driven
by the chemical potential gradient. The velocity of Li-ion movement
is expressed as \citep{mckinnon1983modern}: 
\begin{equation}
\mathbf{v}=-M\nabla\mu,\label{eq:1}
\end{equation}
where $M$ represents the mobility of lithium ions and $\mu$ denotes
the chemical potential. Consequently, the flux of Li ions can be expressed
as: 
\begin{equation}
\mathbf{J}=c\mathbf{v}=-Mc\nabla\mu,\label{eq:2}
\end{equation}
where $c$ is the Li-ion concentration.

In an ideal solid solution, the chemical potential can be written
as \citep{zhang2007numerical}: 
\begin{equation}
\mu=\mu_{0}+RT\ln c-\varOmega_{c}\sigma_{h},\label{eq:3}
\end{equation}
where $\mu_{0}$ is a constant, $R$ is the gas constant, $T$ is the absolute
temperature, which is 298K in this study to simulate isothermal conditions
under room temperature, $X$ is the molar fraction of Li ion, $\varOmega_{c}$
is the partial molar volume of Li ion, and $\sigma_{h}$ is the hydrostatic
stress, given by $\sigma_{h}=(\sigma_{11}+\sigma_{22}+\sigma_{33})\,/\,3$
where $\sigma_{ij}$ are elements in stress tensor. Substituting Eqs.(\ref{eq:1}) and (\ref{eq:3})
into Eq.(\ref{eq:2}): 
\begin{equation}
\mathbf{J}=-D(\nabla c-\frac{c\varOmega_{c}}{RT}\nabla\sigma_{h})\;\mathsf{\mathrm{and}}\;\mathbf{J}\cdot\mathbf{n}=\bar{J},\label{eq:4}
\end{equation}
where $D=MRT$ represents the diffusion coefficient of Li ions in
silicon, $\mathbf{n}$ is the normal vector of the particle boundary, and $\bar{J}$ is the magnitude of the external flux loading. The second equation assumes that the direction of flux is normal to the particle boundary. The conservation of species is expressed as: 
\begin{equation}
\frac{\partial c}{\partial t}+\nabla\cdot\mathbf{J}=0.\label{eq:5}
\end{equation}

\subsection{Diffusion-induced deformation}

For intercalation processes, the lattice constants of the material
are assumed to change linearly with the volume of inserted ions, leading
to the generation of stresses \citep{zhang2007numerical}. Consequently,
the stress-strain relationship for a linear elastic body, incorporating
the chemical strain $\varepsilon_{Li}$ can be expressed as follows
\citep{yang2005interaction}:

\vspace{-3mm}

\begin{subequations}
\begin{equation}
\bm{\varepsilon}_{ij}=\frac{1}{E}[(1+\nu)\bm{\sigma}_{ij}-\nu\bm{\sigma}_{kk}\bm{\mathbf{I}}]+\bm{\varepsilon}_{Li},
\end{equation}
\begin{equation}
\bm{\varepsilon}_{Li}=\frac{\varOmega_{c}(c-c_{0})}{3}\bm{\mathbf{I}},\label{eq:6}
\end{equation}
\end{subequations}
 where $\bm{\varepsilon}_{ij}$ and $\bm{\sigma}_{ij}$ are the strain
and stress components, respectively. $c_{0}$ refers to the reference concentration at the initial stress-free state, which is set to zero in the simulation, as the contribution of pre-existing electrons in the solid is negligible. $E$ is Young's modulus and $\nu$ denotes Poisson's ratio. During
the first cycle, the particle undergoes a phase transformation from
crystalline Si to amorphous $\mathrm{Li_{x}Si}$, resulting in changes to the mechanical properties. So $E$ and $\nu$ are both $(x)$-dependent and defined as follows \citep{shenoy2010elastic}:
\begin{equation}
	E(x)=\frac{m_Ex+n_E}{1+x}[\mathrm{GPa}],\label{eq:E}
\end{equation}
\begin{equation}
	\nu(x)=\frac{m_vx+n_v}{1+x},\label{eq:v}
\end{equation}
where  $x=\frac{x_{max}c}{c_{max}}$, $x_{max}=3.75$ in this study, $c$ is the local Li-ion concentration and $c_{max}$ is the maximum concentration of Li ions in the Si particle. According to the literature, for the crystalline silicon particle, $m_{Ea}=37.96, n_{Ea}=156.13, m_{va}=0.14$ and $n_{va}=0.19$; for the amorphous silicon particle, $m_{Ec}=18.9, n_{Ec}=90.13, m_{vc}=0.24$ and $n_{vc}=0.28$. Also, the Young's modulus degrades during fracture, with the degraded modulus expressed as $E_d=g(d) E $, and $g(d)$ is the degradation function which will be defined in the next section. The expressions for the stress components are as follows: 
\begin{equation}
\bm{\sigma}_{ij}=2\mu(\bm{\varepsilon}_{ij}-\bm{\varepsilon}_{Li})+\lambda(\bm{\varepsilon}_{kk}-\bm{\varepsilon}_{Li})\bm{\mathbf{I}},\label{eq:7}
\end{equation}
where $\mu=\frac{E}{2(1+\nu)}$, $\lambda=\frac{2\nu\mu}{1-2\nu}$ and $\bm{\mathbf{I}}$ is the identical matrix. 

\subsection{Phase field damage}

In the phase field damage model, cracks are represented as interfaces
transitioning smoothly between intact and fully cracked regions \citep{miehe2010thermodynamically}.
This interface is characterised by a scalar $d$, ranging from 0 to 1, where $d=0$ denotes intact area and $d=1$ signifies fully
cracked areas, respectively. Crack propagation is governed by Griffith's
theory, which states that the creation of new crack surfaces releases
strain energy equivalent to the product of the cracked surface area
and the critical energy release rate $G_{c}$, also known as material
toughness. During the phase transition of Si due to Li intercalation,  $G_{c}$ is also dependent on $x$ and defined as \citep{choi2015simple}: 
\begin{equation}
	G_{c}(x)=\begin{cases}
		k_{G1}x^{2}+k_{G2}x+k_{G3} & \mathrm{if}\,0\leq x<1.5\\
		9.7 & \mathrm{if}\,x\geq1.5
	\end{cases}[\mathrm{J/m^{2}}],\label{eq:Gc}
\end{equation}
where, $k_{G1}=1.9, k_{G2}=2.2$ and $k_{G3}=2.1$ are the fitting parameters from the literature. When $x$ is greater than 1.5, the results in the literature suggest that $G_{c}$ reaches a plateau, so 1.5 was chosen as the critical value. The governing equations incorporate stress equilibrium
in the absence of body forces and a phase field evolution law derived
from Griffith's energy balance, employing a hybrid approach \citep{ambati2015review}:
\vspace{-5mm}
\begin{subequations}
\begin{flushleft}
\begin{flalign}
\boldsymbol{\nabla}\cdot[g(d)\boldsymbol{\sigma}] & =\mathbf{0},\label{eq:8a}
\end{flalign}
\begin{equation}
\frac{G_{c}}{l}(d-l^{2}\nabla^{2}d)=2(1-d)\mathcal{H},\label{eq:8b}
\end{equation}
\begin{equation}
\boldsymbol{\sigma}\cdot\mathbf{n}=\mathbf{\bar{t}}\;\mathsf{\mathrm{and}}\;\mathbf{u}=\mathbf{\bar{u}}\;\mathrm{at}\,\partial\varOmega,\label{eq:8c}
\end{equation}
\begin{equation}
\boldsymbol{\nabla}d\cdot\mathbf{n}=0\;\mathrm{and}\;d=0\;\mathrm{at}\,\partial\varOmega,\label{eq:8d}
\end{equation}
\par\end{flushleft}
\end{subequations}
 where $l$ is the phase field length scale, $\partial\varOmega$ is the boundary with the outward normal $\mathbf{n}$, $\mathbf{\bar{t}}$ is the external force vector. For the external surface of the silicon particle, $\mathbf{\bar{t}}$ is set to zero. This condition is physically representative of a single, isolated particle that is not subjected to external mechanical loads. This is a simplification of the current model as the particle is assumed to work in a solid-state electrolyte which will exert an external load on the particle surface. But the present study aims to characterise the intrinsic fracture propensity of the silicon particle itself, driven by its own volumetric changes. A more complex model considering both internal and external effects will be studied in the future. $\mathbf{u}$ is the displacement vector subject to constraints $\mathbf{\bar{u}}$. Non-zero $\mathbf{\bar{u}}$ are not applied to the external, free surface of the particle. Instead, these constraints are used exclusively to enforce the symmetry conditions of 2D axisymmetric modelling. Nodes on the axis of rotational symmetry are constrained to prevent displacement in the radial direction. $g(d)$ is the phase field degradation function \citep{ai2022coupled}: 
\begin{equation}
g(d)=(1-d)^{2}+k,\label{eq:9}
\end{equation}
with $k=10^{-5}$ to avoid singularity at $d=1$. $\mathcal{H}$ is the
local history field of the maximum tensile elastic strain energy $\psi_{0}^{+}(t)$,
$\mathcal{H}=\mathrm{max}\psi_{0}^{+}.$ A volumetric-deviatoric split
\citep{amor2009regularized} is employed to mitigate damage under
compressive loads. The tensile and compressive strain energies are
defined as follows:

\vspace{-5mm}

\begin{subequations}
\begin{align}
\psi_{0}^{+} & =0.5K\bigl \langle \bm{\varepsilon}_{kk}-\bm{\varepsilon}_{Li}\bigr \rangle _{+}^{2}+\mu(\bm{\varepsilon}^{dev}:\bm{\varepsilon}^{dev}),\label{eq:10a}
\end{align}
\begin{equation}
\psi_{0}^{-}=0.5K\bigl \langle \bm{\varepsilon}_{kk}-\bm{\varepsilon}_{Li}\bigr \rangle _{-}^{2},\label{eq:10b}
\end{equation}
\end{subequations}
 where $K$ is the bulk modulus, $K=\lambda+\frac{2}{3}\mu$. $\bm{\varepsilon}^{dev}$
is the deviatoric elastic strain, $\bm{\varepsilon}^{dev}=(\bm{\varepsilon}_{ij}-\bm{\varepsilon}_{Li})-(\bm{\varepsilon}_{kk}-\bm{\varepsilon}_{Li})\bm{\mathbf{I}}$
and the operator $\bigl \langle *\bigr \rangle _{\pm}$is defined as $\bigl \langle x\bigr \rangle=(x\pm\mid x\mid)/2.$

\subsection{Fatigue damage model}

The fatigue damage model is founded on the fatigue degradation function
proposed by Carrara et al. \citep{carrara2020framework}. The material
toughness $G_{c}$ is degraded due to fatigue damage, which is quantified
using a cumulative history variable $\overline{\alpha}$ and a degradation
function $f(\overline{\alpha})$. Consequently, the governing equation
for the phase field is modified to 
\begin{equation}
\frac{G_{d}}{l}(d-l^{2}\nabla^{2}d)-l\nabla d\cdot\nabla G_{d}=2(1-d)\mathcal{H}.\label{eq:11}
\end{equation}
where $G_{d}=f(\overline{\alpha})G_{c}$, the variable $\overline{\alpha}$ is
\begin{equation}
\overline{\alpha}=\int_{0}^{t}\boldsymbol{H}(\alpha\dot{\alpha})\mid\dot{\alpha}\mid d\tau,\label{eq:12}
\end{equation}
with $\tau$ being the pseudo time for integration and $\boldsymbol{H}$
the Heavisid function used to prevent damage accumulation during unloading.
The history variable $\alpha$ is defined as the active part of the
elastic strain energy density, $\alpha=g(d)\psi_{0}^{+},$ with $\dot{\alpha}$
representing its rate. The fatigue degradation function $f(\overline{\alpha})$,
which describes the degradation of material resistance to fracture
during cyclic loading, is defined as follows: 
\begin{flalign}
f(\overline{\alpha}) & =\begin{cases}
1 & \mathrm{if}\ \overline{\alpha}<\alpha_{T},\\
\left(\frac{2\alpha_{T}}{\overline{\alpha}+\alpha_{T}}\right)^{2} & \mathrm{if}\ \overline{\alpha}\geq\alpha_{T},
\end{cases}\label{eq:13}
\end{flalign}
where $\alpha_{T}=G_{c}/(12l)$ is the fatigue crack threshold, above
which fatigue damage begins to accumulate. The history variables $\mathcal{H}$ and
$\overline{\alpha}$ are updated using the following equations

\begin{subequations}
\begin{equation}
\begin{cases}
\mathcal{H}^{n}=\psi_{0}^{+} & \mathrm{if}\ \psi_{0}^{+}>\mathcal{H}^{n-1},\\
\mathcal{H}^{n}=\mathcal{H}^{n-1} & \mathrm{if}\ \psi_{0}^{+}\leq\mathcal{H}^{n-1},
\end{cases}\label{eq:14a}
\end{equation}
\begin{equation}
\begin{cases}
\mathcal{\overline{\alpha}}^{n}=\mathcal{\overline{\alpha}}^{n-1}+\mid\alpha^{n}-\alpha^{n-1}\mid & \mathrm{if}\ \alpha^{n}>\alpha^{n-1},\\
\mathcal{\overline{\alpha}}^{n}=\mathcal{\overline{\alpha}}^{n-1} & \mathrm{if}\ \alpha^{n}\leq\alpha^{n-1},
\end{cases}\label{eq:14b}
\end{equation}
\end{subequations}
 where the superscripts $n$ and $n-1$ correspond to the $n$th time
step and the previous one, respectively.

It is well-established that the significant volumetric expansion of silicon during lithiation constitutes a large deformation process. Consequently, many studies have utilized sophisticated finite strain and elasto-plastic frameworks to capture these effects in detail \citep{zhao2011concurrent,chen2014phase}. However, the primary objective of the present study is to investigate the initiation and propagation of brittle fracture, a process fundamentally governed by the storage and release of elastic strain energy. Also, the computational efficiency of the framework was a prerequisite for conducting an extensive parametric study that forms a core contribution of the work. Therefore, a computationally efficient small strain, linear elastic framework is adopted. The model effectively incorporates the crucial aspect of material evolution and softening via concentration-dependent mechanical properties, serving as a computationally efficient proxy for more complex plastic behaviour.

\section{Finite element implementation}
\label{sec3} 
In the following, we present the finite element (FE) discretisation of the mass transport equation (Section~\ref{Sec:FEmasstransport}),  followed by the discretisation of the deformation (Section~\ref{Sec:FEdef}) and fracture (Section~\ref{Sec:FEPF}) problems, the formulation of the coupled problem (Section~\ref{Sec:SolutionScheme}), and the implementation details of the model (Section~\ref{Sec:implementation}). While the discretisation is performed in COMSOL Multiphysics, the derivations in this section are essential to define the non-standard couplings and the fully coupled stiffness matrix, which constitute our core contribution. This section provides the mathematical blueprint necessary for reproducibility. Common to these sub-sections, the nodal values of the primary fields are interpolated as follows:
\begin{equation}\label{eq:Discretization}
	\mathbf{u}=\sum_{i=1}^n \bm{N}_i \mathbf{u}_i \,,  \hspace{3mm}
	d=\sum_{i=1}^n N_i d_i \,,  \hspace{3mm}
	c=\sum_{i=1}^n N_i c_i
\end{equation}
where Voigt notation has been adopted, $n$ denotes the number of nodes, and $\bm{N}_i$ are the interpolation matrices - diagonal matrices with the nodal shape functions $N_i$ as components. Similarly, the corresponding gradient quantities are discretised as follows,
\begin{equation}\label{eq:Discretization2}
	\bm{\varepsilon}=\sum_{i=1}^n \bm{B}^u_i \mathbf{u}_i \, , \hspace{3mm} 
	\nabla d=\sum_{i=1}^n \mathbf {B}_i d_i\, , \hspace{3mm}  
	\nabla c=\sum_{i=1}^n \mathbf {B}_i c_i\,
\end{equation}
\noindent where $\boldsymbol{B}_{i}^{u}$ denotes the standard strain-displacement
matrices, $B_{i}^{u}=\left[\begin{array}{cc}
	\frac{\partial N_{i}}{\partial x} & 0\\
	0 & \frac{\partial N_{i}}{\partial y}\\
	\frac{\partial N_{i}}{\partial y} & \frac{\partial N_{i}}{\partial x}
\end{array}\right]$, and $B_{i}$ are vectors with the spatial derivatives of the shape
functions, $B_{i}=\left[\begin{array}{cc}
	\frac{\partial N_{i}}{\partial x}, & \frac{\partial N_{i}}{\partial y}\end{array}\right]^{\mathrm{T}}$, and the components of the stress tensor  are
\begin{equation}
	\bm{\sigma}=\sum_{i=1}^n\mathbf{C}_{0}\left[\mathbf{B}_{i}\mathbf{u}_{i}-N_{i}c_{i}\frac{\varOmega}{3}\mathbf{I}^{c}\right], \label{eq:17}
\end{equation}
where $\mathbf{C}_{0}$ is the linear elastic stiffness matrix and
$\mathbf{I}^{c}=\left[1,1,0\right]^{\mathrm{T}}.$

\subsection{FE discretisation of the mass transport problem}
\label{Sec:FEmasstransport}

The weak form of the mass transport problem can be formulated as
\begin{equation}
	\int_{\varOmega}\nabla(\delta c)\cdot\mathbf{J}-\delta c\frac{\mathrm{d}c}{\mathrm{d}t}\mathrm{d}V-\int_{\partial\varOmega}\delta c\overline{J}\mathrm{d}S=0. \label{eq:18}
\end{equation}
The residual vector for the mass transport problem is obtained by discretising Eq.~(\ref{eq:18}), where $\delta c$ represents an arbitrary variation of the Li-ion concentration,
\begin{equation} \label{eq:residualC}
	r_{i}^{c}=\int_{\varOmega}(\mathbf{B}_{i}^{\mathrm{T}}D\nabla c-\mathbf{B}_{i}^{\mathrm{T}}\frac{D\varOmega_{c}c}{RT}\nabla\sigma_{h}+N_{i}\frac{\mathrm{d}c}{\mathrm{d}t})\mathrm{d}V+\int_{\partial\varOmega}N_{i}\overline{J}\mathrm{d}S.
\end{equation}
A diffusivity matrix can then be defined,
\begin{equation}
	K_{ij}^{c}=\frac{\partial r_{i}^{c}}{\partial c}=\int_{\varOmega}(\mathbf{B}_{i}^{\mathrm{T}}D\mathbf{B}_{j}-\mathbf{B}_{i}^{\mathrm{T}}\frac{D\varOmega_{c}}{RT}N_{j}\mathbf{B}_{j}\sigma_{h})\mathrm{d}V , \label{eq:Kc}
\end{equation}
where the subscripts $i$ and $j$ for the bold variables denote contributions from nodes $i$ and $j$ rather than matrix components. Considering the coupling between mass transport and deformation, and the coupling between mass transport and phase field:
\begin{equation}
	K_{ij}^{c\mathbf{u}}=\frac{\partial r_{i}^{c}}{\partial \mathbf{u}}=-\frac{D\varOmega_{c}K}{RT}\int_{\varOmega}\mathbf{B}_{i}^{\mathrm{T}}\cdot\nabla\mathbf{B}_{j}\mathrm{d}V , \label{eq:Kcu}
\end{equation}
\begin{equation}
	K_{ij}^{cd}=\frac{\partial r_{i}^{c}}{\partial d}=-\frac{2ED\varOmega_{c}}{3RT}\int_{\varOmega}\mathbf{B}_{i}^{\mathrm{T}}\nabla[(1-d)]{N_j}\mathrm{tr}(\mathbf{C}_{0}\mathbf{\epsilon})\mathrm{d}V , \label{eq:Kcd}
\end{equation}
Together with a concentration capacity matrix,
\begin{equation} \label{eq:concentrationM}
	M_{ij} = \int_\Omega  {N_i^T{N_j}\mathrm{d}V},
\end{equation}
and a diffusion flux vector,
\begin{equation} \label{eq:diffusionFlux}
	f_i^c = \int_{\partial\mathrm{\Omega}_J} N_i^T \bar{J} \, \mathrm{d}S.
\end{equation}
Subsequently, the discretised moisture transport equation reads,
\begin{equation} \label{eq:moisturesystem}
	\bm{K}^c \mathbf{c} + \bm{M} \dot{\mathbf{c}} = \mathbf{f}. 
\end{equation}

\subsection{FE discretisation of the deformation problem}
\label{Sec:FEdef}

The weak form of the deformation problem can be formulated as
\begin{equation}\label{eq:weakformdef}
	\int_{\Omega} \left[g(d)  \bm{\sigma} : \delta \bm{\varepsilon} \right]  \, \mathrm{d}V-\int_{\partial\varOmega}\bar{\mathbf{t}}\cdot\delta\mathbf{u}\mathrm{d}S = 0. 
\end{equation}
Making use of the FE discretisation given in Eqs.~(\ref{eq:Discretization}) and (\ref{eq:Discretization2}), and considering that Eq.~(\ref{eq:weakformdef}) must hold for arbitrary values of the primal field variables, the residual can be derived as follows: 
\begin{equation} \label{eq:residualStagU}
	\mathbf{r}_{i}^{\mathbf{u}}=\int_{\varOmega}g(d)(\mathbf{B}_{i}^{\mathbf{u}})^{\mathrm{T}}\boldsymbol{\sigma}\mathrm{d}V-\int_{\partial\varOmega}N_{i}\bar{\mathbf{t}}\mathrm{d}S, 
\end{equation}
 The components of the stiffness matrices are derived by differentiating the residual to the incremental nodal variable, as follows:
\begin{equation}\label{Eq:Ku}
	K_{ij}^{\mathbf{u}} = \frac{\partial \mathbf{r}_{i}^{\mathbf{u}}}{\partial \mathbf{u}_{j}} = \int_\Omega \left\{ g(d) {(\mathbf{B}_i^{\mathbf{u}})}^T \mathcal{c} \mathbf{B}_j^{\mathbf{u}} \right\} \, \mathrm{d}V  \,, 
\end{equation}
and considering the coupling between deformation and phase field  and the coupling between deformation and mass transport, there are
\begin{equation}\label{Eq:Kud}
	K_{ij}^{\mathbf{u}d} = \frac{\partial \mathbf{r}_{i}^{\mathbf{u}}}{\partial d_{j}} = -2\int_\Omega \left\{ (1-d)N_j {(\mathbf{B}_i^{\mathbf{u}})}^T \mathbf{C}_{0} \mathbf{B}_j^{\mathbf{u}}\mathbf{u}_j \right\} \, \mathrm{d}V  \,, 
\end{equation}
\begin{equation}
	K_{ij}^{\mathbf{u}c}=\frac{\partial\mathbf{r}_{i}^{u}}{\partial\mathbf{c}_{j}}=\int_{\varOmega}(1-d)^2(\mathbf{B}_{i}^{u})^{\mathrm{T}}\frac{\partial\mathbf{C}}{\partial c_j}\mathbf{B}_{k}^{\mathbf{u}}\mathbf{u}_k N_j\mathrm{d}V, \label{eq:Kuc}
\end{equation}
where $\mathbf{C}$ is the elastic stiffness tensor.
Together with a displacement vector,
\begin{equation}
	f_{i}^{\mathbf{u}}=-\int_{\partial\varOmega_{\mathbf{u}}}N_{i}\bar{\mathbf{t}}\mathrm{d}S \label{eq:30}
\end{equation}

\subsection{FE discretisation of the phase-field problem}
\label{Sec:FEPF}
The weak form of the phase-field problem can be formulated as
\begin{equation}\label{eq:weakformPF}
\int_{\Omega} \left[-2(1-d)\delta d \, \mathcal{H}   + G_c \left( \frac{d}{\ell} \delta d
	+ \ell \nabla d \cdot \nabla \delta d \right) \right]  \, \mathrm{d}V
\end{equation}
The residue can be expressed as follows
\begin{equation}
	r_{i}^{d}= \int_\Omega \left[ -2(1-d) N_{i}  \left. \mathcal{H} \right. + G_c \left(\dfrac{d}{\ell} N_{i}  + \ell \mathbf{B}_{i}^T \nabla d \right) \right] \, \mathrm{d}V, \label{eq:ResidualPF}
\end{equation}
The component of the stiffness matrices is derived by differentiating the residual with respect to the incremental nodal variable, as follows
\begin{equation}
	K_{ij}^d = \dfrac{\partial r_{i}^d}{\partial d_{j}} = \int_\Omega \left\{ \left[ 2 \mathcal{H} + \frac{G_c}{\ell} \right]  N_{i} N_{j} + G_c \ell \mathbf{B}_i^T \mathbf{B}_j  \right\} \, \mathrm{d}V  \, ,\label{eq:Kd}
\end{equation}
and considering the coupling between phase field and mass transport,  and the coupling between phase field and deformation, there are
\begin{equation}
	K_{ij}^{d\mathbf{u}} = \dfrac{\partial r_{i}^d}{\partial \mathbf{u}_{j}} = -2\int_\Omega  (1-d)N_i(\mathbf{B}_j^\mathbf{u})^T \mathbf{C}_0 \mathbf{B}_k^\mathbf{u}\mathbf{u}_k \mathrm{d}V ,\label{eq:Kdu}
\end{equation}
\begin{equation}
	K_{ij}^{dc} = \dfrac{\partial r_{i}^d}{\partial c} = -2\int_\Omega  (1-d)N_i(\frac{1}{2}\epsilon: \dfrac{\partial \mathbf{C}}{\partial c_j}:\epsilon)\mathrm{d}V+\int_\Omega\dfrac{\partial G_c}{\partial c_j}N_j(\frac{d}{l}N_i+l\mathbf{B}_i^T\nabla d) \mathrm{d}V ,\label{eq:Kdc}
\end{equation}
\subsection{Coupled scheme}
\label{Sec:SolutionScheme}

\begin{figure}[!ht ]
	\centering{}\includegraphics[width=13cm]{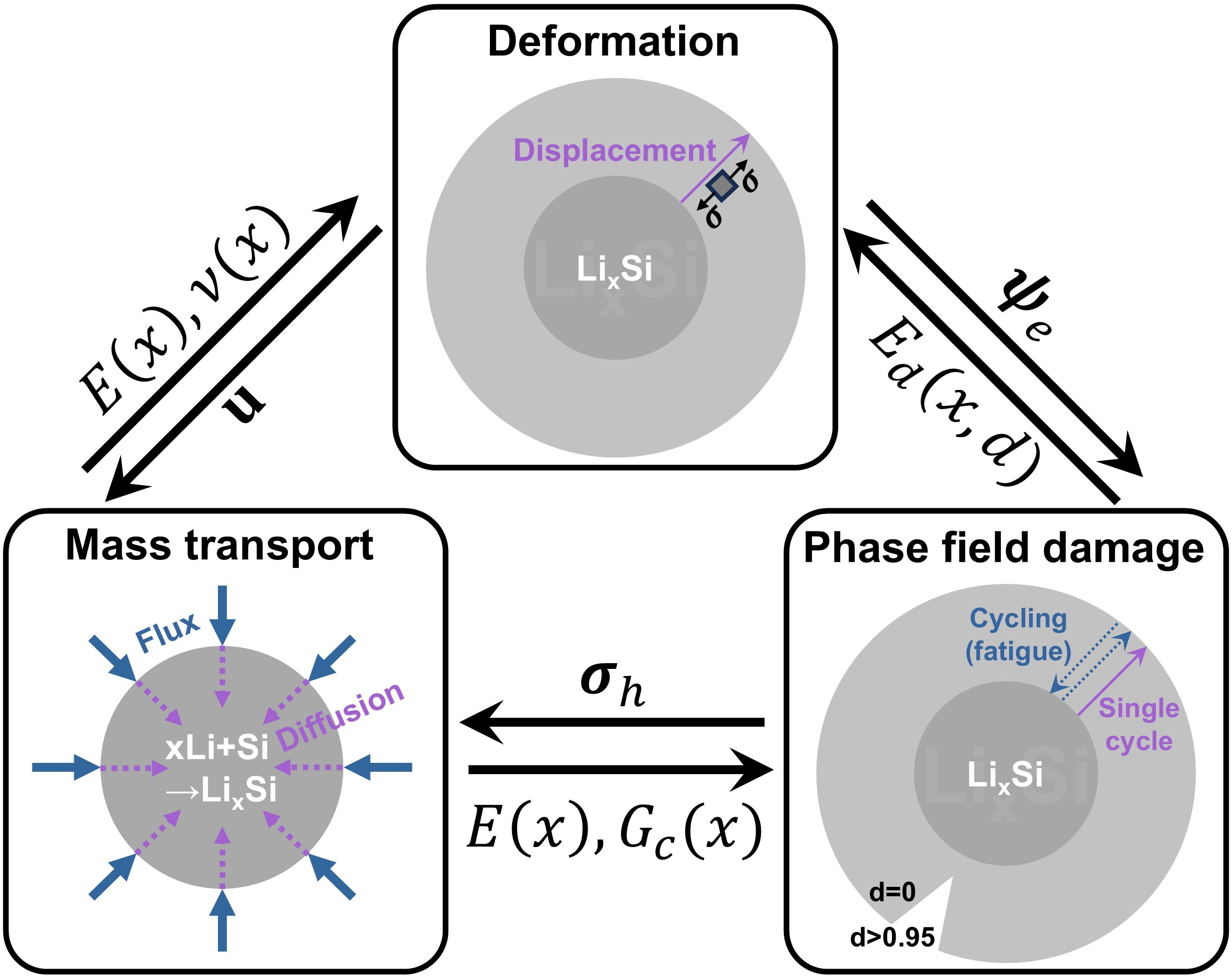}
	\centering{}\caption{Coupling relationships between multiple physics.}
	\label{fig:1} 
\end{figure}

The deformation, diffusion, and phase field fracture problems are fully coupled through a series of variables, as illustrated in Fig. \ref{fig:1}. First, mass transport influences the stress field in the fracture process zone through diffusion-induced chemical strain. Additionally, the diffusion of Li-ions in the Si particle induces the phase transformation of $\mathrm{Li_{x}Si}$, altering the phase-dependent Young's modulus ($E(x)$) and Poisson's ratio ($\nu(x)$), and further influencing the evolution of the phase field through phase-dependent toughness ($G_c(x)$). Second, the deformation of the particle ($\mathbf{u}$) inversely impacts the diffusion of Li-ions within the Si particle, while the resulting mechanical stress ($\bm{\sigma}_{ij}$) governs the evolution of the phase field. Finally, phase field fracture impedes the diffusion of Li-ions ($D(d)$) and degrades the Young's modulus ($E_d(x, d)$). The linearised finite element system,
\begin{equation} \label{eq:coupledScheme}
	\left[\begin{array}{ccc}
		\boldsymbol{K}^{\mathbf{u}} & \boldsymbol{K}^{\mathbf{u}d} & \boldsymbol{K}^{\mathbf{u}c}\\
		\boldsymbol{K}^{d\mathbf{u}} & \boldsymbol{K}^{d} & \boldsymbol{K}^{dc}\\
		\boldsymbol{K}^{c\mathbf{u}} & \boldsymbol{K}^{cd} & \boldsymbol{K}^{c}
	\end{array}\right]\left[\begin{array}{c}
		\mathbf{u}\\
		\boldsymbol{d}\\
		\boldsymbol{c}
	\end{array}\right]+\left[\begin{array}{ccc}
		0 & 0 & 0\\
		0 & 0 & 0\\
		0 & 0 & \boldsymbol{M}
	\end{array}\right]\left[\begin{array}{c}
		\dot{\mathbf{u}}\\
		\boldsymbol{\dot{d}}\\
		\boldsymbol{\dot{c}}
	\end{array}\right]+\left[\begin{array}{c}
		\boldsymbol{f}^{\mathbf{u}}\\
		0\\
		\boldsymbol{f}^{c}
	\end{array}\right]=\left[\begin{array}{c}
		\boldsymbol{r}^{\mathbf{u}}\\
		\boldsymbol{r}^{d}\\
		\boldsymbol{r}^{c}
	\end{array}\right] 
\end{equation} 
is solved incrementally using the Newton-Raphson method. For a single cycle, the solution scheme adopts a \emph{staggered} approach \cite{miehe2010phase}, where the displacement, Li-ion concentration, and phase field problems are solved sequentially to reduce computational load. In contrast, the fatigue solution scheme utilises a \emph{fully-coupled} approach \citep{kristensen2020phase}, where the three variables are computed simultaneously. This is because, for the fatigue fracture problem, the staggered approach necessitates a significantly larger number of load increments to achieve equilibrium.

\subsection{Model implementation}
\label{Sec:implementation}

\begin{figure}[!ht ]
	\centering{}\includegraphics[width=13cm]{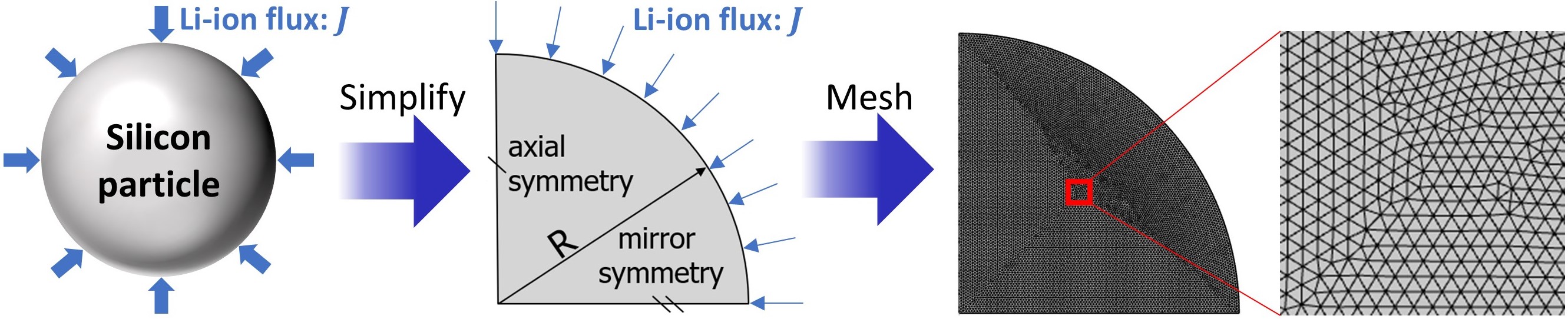}
	\centering{}\caption{Geometry and meshing in the simulation.}
	\label{fig:2} 
\end{figure}

The implementation is carried out in the finite element package COMSOL Multiphysics. Within COMSOL, the Solid Mechanics, Transport of Diluted Species, and Helmholtz Equation modules are selected for this study. Combining Solid Mechanics, Transport of Diluted Species, and Helmholtz Equation modules is a validated methodology in recent literature. Notably, Ai, et al. \citep{ai2022coupled} employed these modules to address a similar problem regarding battery NMC particle fatigue. As shown in Fig. \ref{fig:2}, for simplicity, a 2D axisymmetric model
is constructed to represent the spherical Si particles. The geometry
is meshed using free triangular elements, with the minimum and maximum
element sizes controlled at $r_{p}/200$ and $r_{p}/100$, respectively,
to ensure good mesh quality, where $r_{p}$ is the particle radius.
The partial differential equations are solved using the PARDISO solver
and an implicit Backward Differentiation Formula (BDF). The modules
are solved in a segregated manner. This research utilised Queen Mary's Apocrita HPC facility, supported by QMUL Research-IT \cite{king_2017_438045}. 

The lithium ion flux into or out of the particle is $s(t)\bar{\mathbf{J}}$
where $s(t)$ is a sign function used to control the charge, relaxation, and discharge states. The transition between these states is smoothed
to facilitate numerical convergence, with the transition state occupying
0.02 of each cycle period. The magnitude of the flux $\bar{J}$
is written as 
\begin{equation}
	\bar{J}=c_{\mathrm{max}}\mathrm{\frac{volume}{area}}\frac{C}{3600}\:[\mathrm{mol\,m^{-2}s^{-1}}],\label{eq:32}
\end{equation}
where $C$ is the C-rate which is the inverse of the time (in hours)
required to fully charge or discharge a battery. Once
the lithium concentration reaches the cut-off values $c=0$ or
the maximum concentration, it remains constant at these limits. 
In this study, the fully cracked threshold of the phase field is set
to 0.95, meaning that areas where $d\geq0.95$ are considered cracked.
Within the assumed solid-state electrolyte system, a crack forms a physical void that breaks the ionic transport pathway. Once this contact is lost, the fractured region becomes electrochemically inactive. To model this phenomenon, the diffusion coefficient, $D$, is defined as dependent phase field damage $d$: 
\begin{equation}
	D(d)=\begin{cases}
		D_{0} & \mathrm{if}\,d<0.95,\\
		0 & \mathrm{if}\,d\geq0.95,
	\end{cases}[m^{2}s^{-1}]\label{eq:36}
\end{equation}
where $D_{0}=3\times10^{-6}\mathrm{m^{2}s^{-1}}$ is the Li-ion diffusion
coefficient in Si. For the pre-existing crack, it is introduced by setting
up a non-zero history field $\mathcal{H}$ to a local crack partition
domain at $t=0$, using \citep{ai2022coupled}
\begin{equation}
	\mathcal{H}=\alpha_{0}\mathrm{exp}\left(-\frac{100s^{2}}{l^{2}}\right),\label{eq:37}
\end{equation}
where $\alpha_{0}=10^{12}\mathrm{Jm^{-3}}$, $s$ is the distance
to the crack plane and $l=0.01\mathrm{\mu m}$ is the phase field
length scale \citep{klinsmann2016modeling}. The width of pre-existing
crack $w_{cr}$is set to a small value relative to the particle radius
$r_{p}$, $w_{cr}=r_{p}/50$. The particle varies as it is a study
objective in this research. The impact of the solid electrolyte interphase
(SEI) formation on newly created crack faces is disregarded because
of two reasons: first, the electrolyte is assumed to be solid according
to the experimental setting of the referred experimental study \citep{liu2012size}, which can largely reduce the SEI formation;
second, the flux loading magnitude remains relatively stable at high
Coulombic efficiencies (99.99\%), and the SEI layer, being porous
and weak, does not compromise the structural integrity of electrode
particles \citep{ai2022coupled}. Other material properties and model
parameters are summarised in Table. \ref{table: 1}. 
\vspace{-5mm}
\begin{table}[h]
	\centering{}\caption{Material properties and model parameters in the study.}
	\label{table: 1}%
	\begin{tabular}{ccccc}
		\hline 
		Symbol  & Description  & Value  & Unit  & Source\tabularnewline
		\hline 
		$\varOmega_{c}$  & Partial molar volume of Li  & $8.5\times10^{-6}$  & $\mathrm{m^{3}mol^{-1}}$  & \citep{mesgarnejad2019phase}\tabularnewline
		$c_{\mathrm{max}}$  & Maximum lithium concentration  & $3.11\times10^{5}$  & $\mathrm{mol\,m^{-3}}$  & \citep{liu2020cracks}\tabularnewline
		$D_{0}$  & Li-ion diffusion coefficient in $\mathrm{Li_{x}Si}$  & $3\times10^{-6}$  & $\mathrm{m^{2}s^{-1}}$  & \citep{wang2017multiphysics}\tabularnewline
		\hline 
	\end{tabular}
\end{table}

\section{Results and discussion}

\label{sec4} 

\subsection{Model validation}

\begin{figure}[!ht ]
\centering{}\includegraphics[width=13cm]{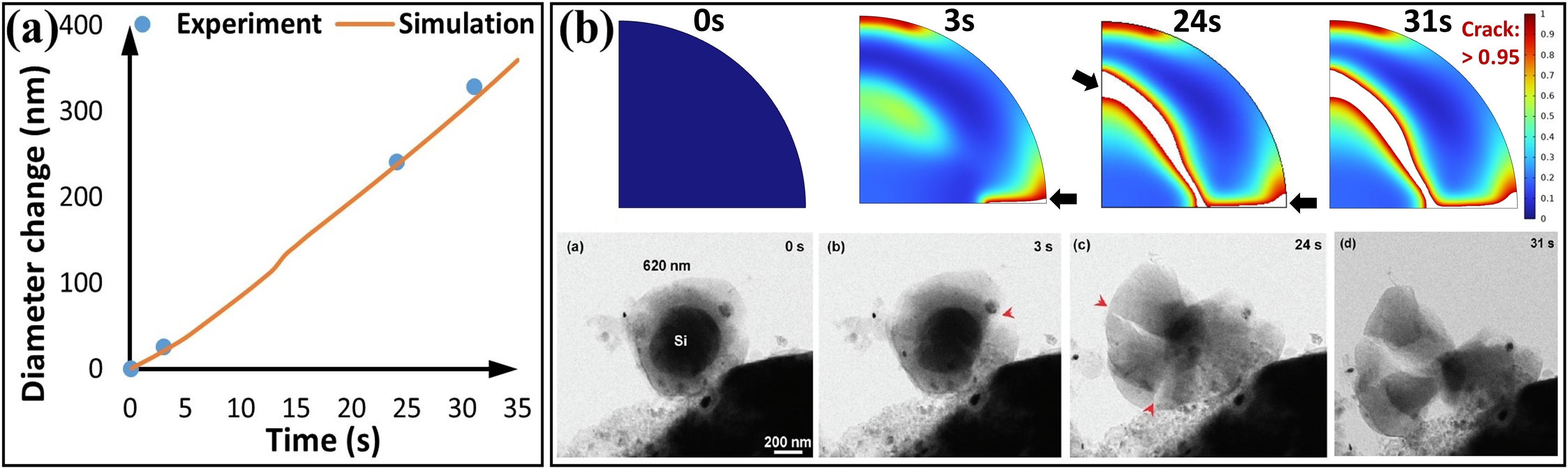} \centering{}\caption{Validation of simulation results with experimental results \citep{liu2012size}:
(a) change of particle diameter and (b) cracking and fracturing behaviours. The value '1' in the legend means the maximum value of phase field damage, and value '0.95' is defined the critical value of crack.}
\label{fig:3} 
\end{figure}

The multiphysics model can be validated by the experimental results
reported by Liu, et al. \citep{liu2012size}. In that study, the cracking
behaviour of a Si particle with a diameter of 620 nm was recorded. The literature  shows an incomplete lithiation at 49s, so a 60C rate was selected to approximate experimental conditions.
The simulation results align with the experimental findings in terms
of diameter change, crack initiation
time and position, the locations of other cracks, and the complete
fracturing time of the particle when the crack propagates throughout,
as shown in Fig. \ref{fig:3}(a) and (b). The diameter change in the simulation was obtained by probing the displacement of a point on the uncracked area of the particle surface. At 3s, the crack initiates
from the surface in both experiments and simulations; at 24s, the initial
crack continues propagating, and an internal
crack shows in both experiments and simulations; at 31s, the crack
penetrates the Si particle, leading to the simultaneous fracture of
the particle. Moreover, while the Si particle in the experiment is crystalline, the particle expansion is nearly isotropic, demonstrating that our simplified 2D axisymmetric framework with isotropic material properties is enough to simulate the particle growth in the experiment.

\subsection{Cracking and fracture mechanism of Si particle}

\begin{figure}[!ht]
\centering{}\includegraphics[width=13cm]{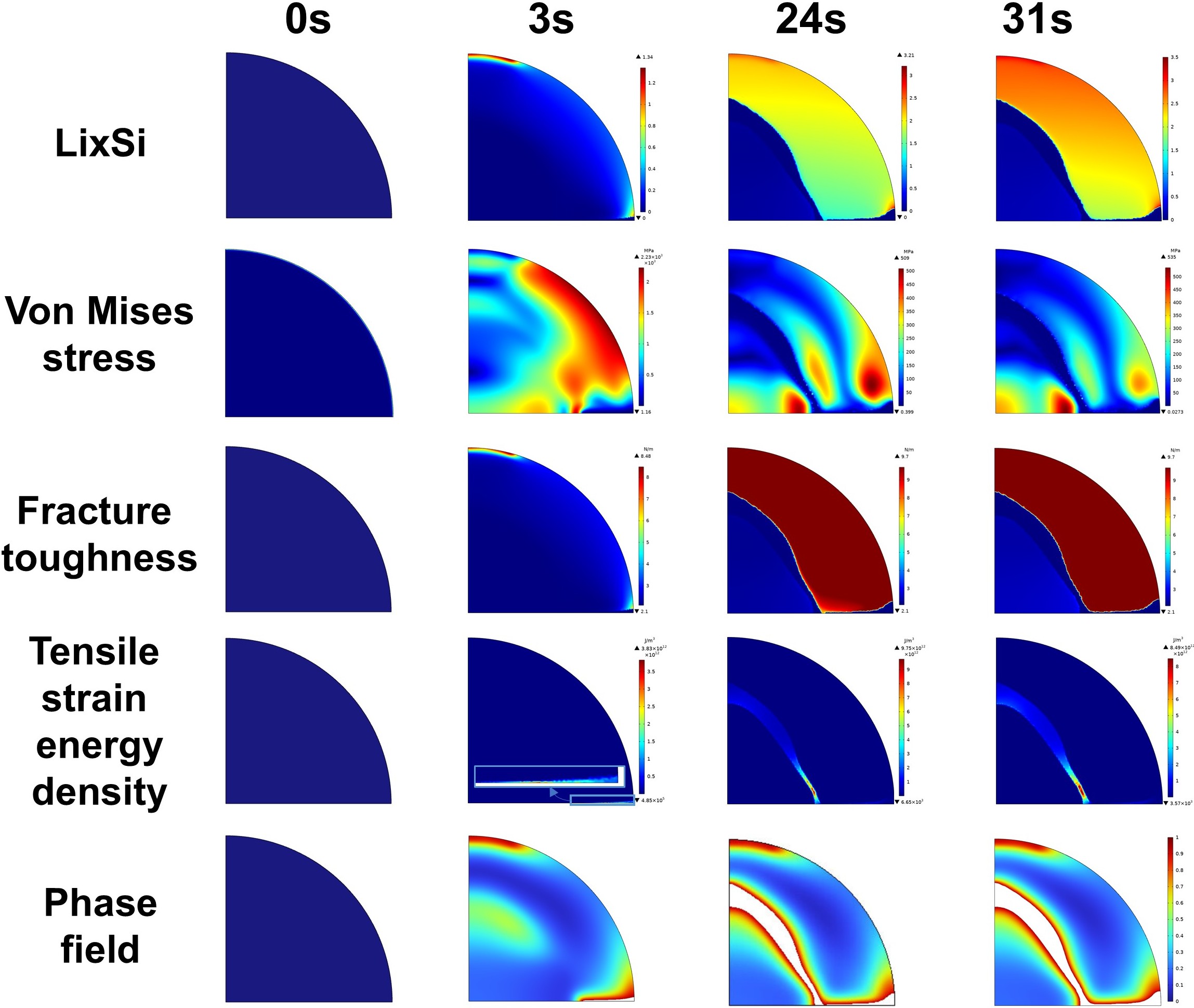} \centering{}\caption{Cracking and fracture mechanism of Si particle.}
\label{fig:4} 
\end{figure}

As shown in Fig. \ref{fig:4}, when Li ions diffuse into the Si particle,
the diffusion coefficient within the particle is considerably lower
than that in the surrounding environment. Such concentration disparity causes increased tensile strain energy, large local stress difference and inhomogeneous fracture toughness, leading to crack initiation. Then, the cracks hinder the Li-ion diffusion which exacerbates the concentration disparity and leads to stress concentrating ahead of the crack tip. The stress concentration, combined with high tensile strain energy and relatively low fracture toughness, facilitates the crack propagation and eventually particle fracture. This accumulation amplifies
the local stress at the crack tip, promoting crack propagation and
eventually causing the particle to fracture. The Von Mises stress
distribution reveals a high-stress region near the particle surface,
despite the absence of visible cracks. This phenomenon arises due
to the increase in fracture toughness with Li-ion concentration (denoted
as x in $\mathrm{Li_{x}Si}$). At this stage, the surface region exhibits
enhanced fracture toughness, enabling it to sustain elevated stress
levels. Note that some studies observed a surface hoop stress reversal when considering plasticity \citep{liu2012size,cui2012finite}, which may trigger the surface crack. This phenomenon is not observed in the current study, so the plasticity will be incorporated in the model in future studies.

\subsection{Effects of particle diameter and charge rate}

Particle diameter and charge rate influence the Li-ion diffusion spatially and temporally, respectively, and change the Li-ion concentration distribution and stress distribution, thus affecting cracking and fracture behaviours of the Si particle. In this section, the fracture behaviours and damage ratios of different particle diameters and charge rates are compared, and a contour map of cracking and fracturing behaviours is generated.

\textbf{Cracking and fracture behaviours} 

\begin{figure}[!ht]
\centering{}\includegraphics[width=13cm]{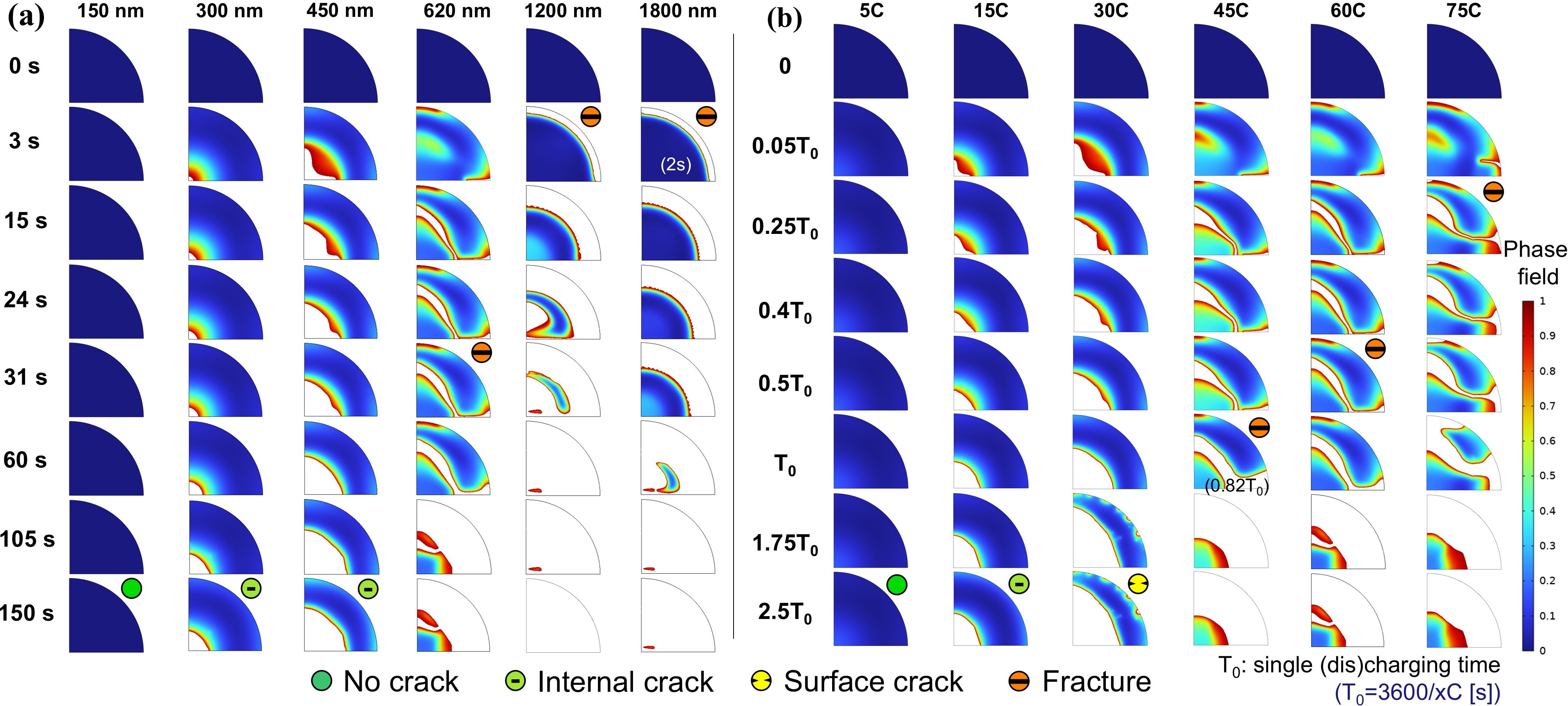}
\centering{}\caption{Cracking and fracture behaviours of particle (a) with different diameters
(@60C of charge rate) and (b) under different charge rates (@620 nm
of particle diameter).}
\label{fig:5} 
\end{figure}

In Fig. \ref{fig:5}(a), for particles with diameters below 150 nm, lithium (Li) ions diffuse uniformly throughout the entire particle, resulting in a homogeneous Li-ion concentration and negligible chemical stress, thereby preventing crack formation. For particles with diameters ranging from 150 nm to 450 nm, Li ions diffuse through most of the particle but not the central region, causing significant swelling in the outer regions and tensile stress in the core, leading to central cracking. In particles with diameters between 450 and 1200 nm, Li ions accumulate at the boundary, creating a pronounced concentration gradient that initiates cracking. As Li ions penetrate further into the particle, swelling in the outer regions generates tensile stress within the internal structure, resulting in cracks at the particle centre. For particles with diameters of 1200 nm or larger, Li ions densely accumulate at the boundary, causing complete boundary cracking. Additionally, particles with diameters exceeding 620 nm begin to fracture, with some undergoing complete fracture. The time points of fracture and complete fracture, as indicated in the figure, show that as the diameter increases, the time to fracture decreases while the time to complete fracture increases. Note that the 'complete fracture' in this study means a continuous crack path (a collection of locally failed points) fully traverses the particle, leading to its fragmentation, which is the analogue of the physical pulverisation observed experimentally in large silicon particles. Notably, particles with a diameter of 1800 nm do not fracture by the end of the process. This behaviour will be further elucidated in the analysis of normalised crack volume.

In Fig. \ref{fig:5}(b), the charging rate demonstrates effects analogous to those observed with particle diameter. Note that the charge rate in the simulation is up to 75C, because high C-rate simulations are essential for mapping the full chemo-mechanical failure envelope and represent the realistic localized currents particles experience even during moderate cell charging \citep{fan2013mechanical,zhu2019fast}. As shown in the figure, at low charging rates (C-rate < 5C), the influx of Li ions into the particle balances their diffusion out of the boundary, preventing accumulation and ensuring uniform stress distribution. At moderate charging rates (5C < C-rate < 15C), Li ions penetrate most of the particle except the central region, leading to swelling in the outer regions and tensile stress in the core, which ultimately results in cracking. At high charging rates, Li ions rapidly accumulate at the particle boundary, creating substantial chemical stress that causes surface cracking. Furthermore, particles begin to fracture at charge rates above 45C. As the charge rate increases, the times to fracture and complete fracture decrease. Note that in the study of Liu et al. \citep{liu2012size}, they concluded that lithiation rate was insignificant under their specific experimental conditions, which seems to conflict with our simulation results. At sufficiently high charging rates, the external Li-ion flux can exceed the maximum velocity of the interfacial reaction front. Under these conditions, the system is forced into a diffusion-limited regime, where Li-ions accumulate in the outer shell, creating a steep concentration gradient \citep{chen2020molecular}. At this high-rate limit, the stress generation and fracture behaviour of c-Si will also become strongly rate-dependent, converging with the predictions of our model. Also, an in-situ experimental study shows that a low charge rate only causes electrode expansion, while a high charge rate can lead to electrode fracture and pulverisation \citep{dienwiebel2022visualization}. Some other studies \citep{liu2020cracks,choi2023phase} also showed the rate-dependent behaviours of Si particle fracture. It is well-established from experimental observations that the initial lithiation of crystalline silicon proceeds via a two-phase mechanism, characterised by a sharp, reaction-limited interface moving between the pristine c-Si core and an amorphous a-$\mathrm{Li_{x}Si}$ shell. In this study, although a continuum framework, rather than an explicit sharp-interface or phase-field phase transformation model, was used, by defining the mechanical properties ($E$ and $\nu$) and fracture toughness ($G_{c}$) as continuous functions of the local lithium concentration, a steep gradient in mechanical properties and chemical strain across the lithiation front can be observed, especially for simulations with large particle diameter or high charge rate.

\textbf{Damage ratio}

\begin{figure}[!ht]
\centering{}\includegraphics[width=13cm]{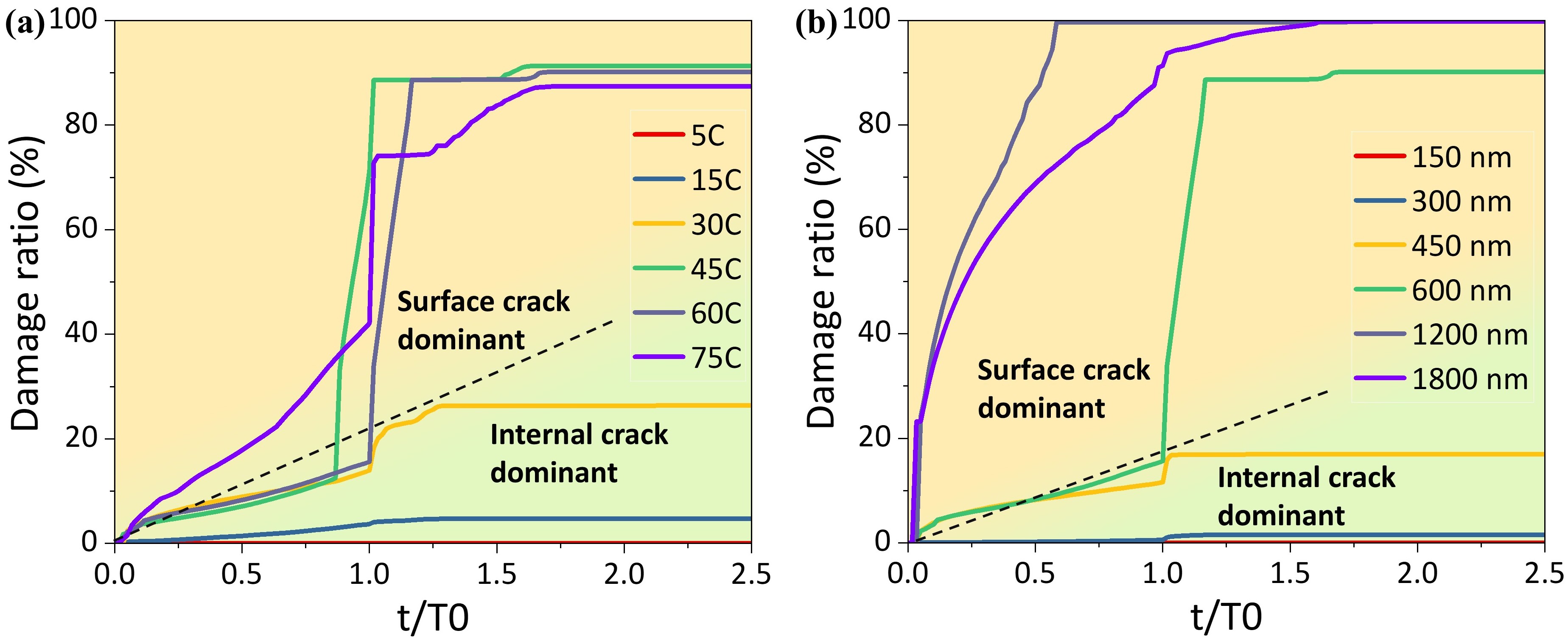}
\centering{}\caption{Damage ratios of particles (a) with different diameters (@60C
of charge rate) and (b) under different charge rates (@620nm of particle
diameter).}
\label{fig:6} 
\end{figure}

Damage ratio is defined as the ratio of cracked particle volume to total particle volume. Fig. \ref{fig:6}(a) and (b) show the damage ratios of the particles with the aforementioned parameters. By analysing
the damage ratio as a function of the cycling state, cracking
behaviours can be classified into two categories: (i) internal cracks dominant, and (ii) surface crack dominating. In the first category, the cracking rate is the
low, resulting in a small final damage ratio during the first
cycle, and the particle remains intact. In the second category,
the cracking rate is high, ultimately resulting in particle fracture. Furthermore, the results reveal that surface cracks
cause more severe damage compared to internal cracks. The influence
of pre-existing cracks (notches) will be examined in a subsequent chapter to
provide further clarification.

\textbf{Contour map of cracking and fracturing behaviours}

\begin{figure}[!ht]
\centering{}\includegraphics[width=13cm]{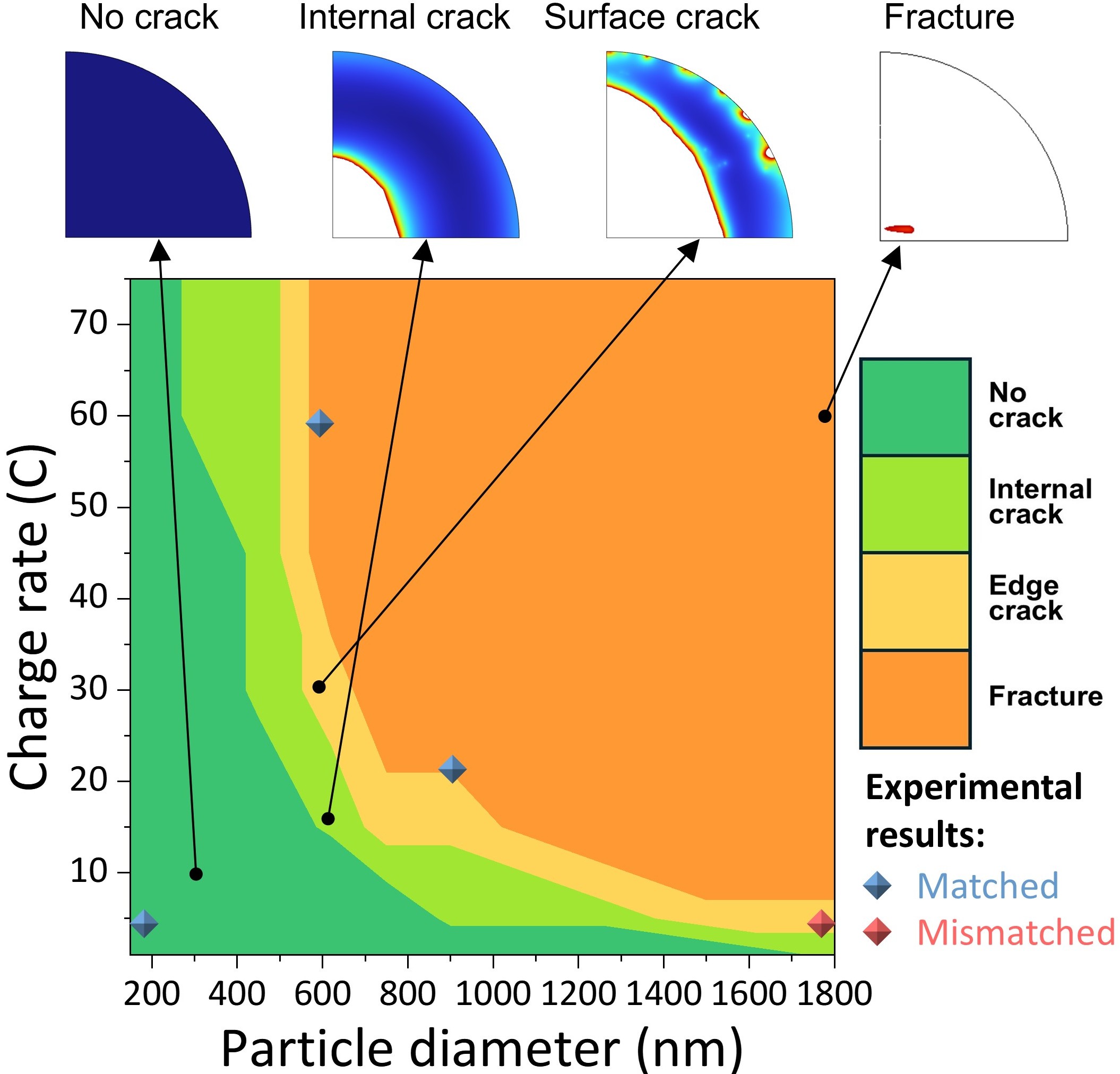}
\centering{}\caption{Contour map of cracking and fracturing behaviours.}
\label{fig:7} 
\end{figure}

To systematically
characterise the fracturing behaviours under diverse conditions, a
contour map (Fig. \ref{fig:7}) was constructed to illustrate the
cracking and fracturing patterns of Si particles during the first
cycle. The map was derived from over 60 simulations, encompassing
a range of charge rates and particle diameters, and classified four
distinct fracturing behaviours---no crack, internal crack, surface
crack, and fracture---based on the particle's final state after the
first cycle. Cracks were absent in particles with diameters below
150 nm. Fracturing was observed in particles larger than 450 nm when
subjected to high charge rates (>60C). Additionally, particles larger
than 1200 nm exhibited fracturing at charge rates exceeding 10C. Experimental
results \citep{liu2012size} plotted on the figure demonstrated strong
agreement with simulation predictions, except for particles with a
diameter of 1800 nm. This discrepancy may be attributed to the presence
of pre-existing micro-cracks in larger particles, which exacerbated
fracturing beyond the predicted behaviour. This finding suggests that
pre-existing notches significantly influence the cracking and fracturing
behaviours of Si particles. Consequently, the influence of pre-existing
cracks is investigated in the following section.

\subsection{Effects of pre-existing notches}

Micro-cracks in the Si particle generated during production are common. And in this section, the effects of pre-existing notches on the cracking and fracture behaviours of Si particle under different charges rates and with different locations, lengths and orientations are investigated.

\textbf{Different charge rates and notch locations} 

\begin{figure}[!ht]
\centering{}\includegraphics[width=13cm]{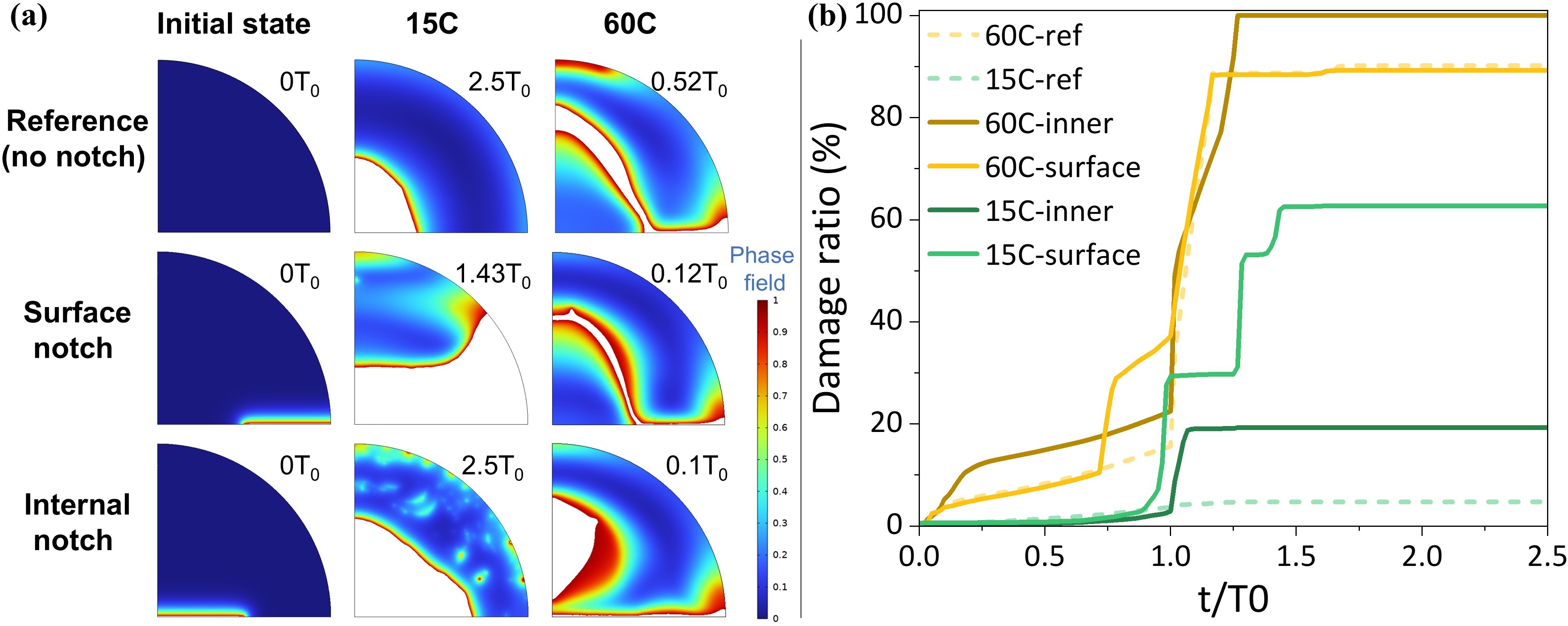}
\centering{}\caption{(a) Fracture time and (b) normalised crack volumes
of particles subjected to varying charge rates and different types
of pre-existing notches. The particle diameter is 620 nm.}
\label{fig:8} 
\end{figure}

Fig.\ref{fig:8}(a) depicts
the cracking and fracturing behaviours of particles subjected to varying
charge rates and different types of pre-existing notches. Each charge
rate corresponds to a characteristic cracking mode: (i) internal cracks dominant, and (ii) surface cracks dominant. In both cases, the pre-existing notch accelerates the particle fracture and enlarges the damage ratio.   
In the first cracking mode (e.g., 15C charge rate), internal pre-existing
cracks have minimal influence on particle fracturing, whereas surface
pre-existing notches significantly accelerate the fracturing process.
In the second cracking mode (e.g., 45C charge rate), both surface
and internal pre-existing notches reduce the time to particle fracturing,
with internal cracks exerting a more pronounced effect by accelerating
the fracturing process. 

To further illustrate the effects of pre-existing notches on different
cracking modes, the normalised crack volume as a function of the cycling
state is presented in Fig. \ref{fig:8}(b). For the cracking mode
characterised by internal cracks dominating (e.g., 15C), the presence of
a surface pre-existing crack accelerates the cracking process, resulting
in a larger damage ratio. In contrast, for cracking modes dominated by surface cracks (e.g., 60C), internal pre-existing
cracks significantly increase the cracking speed.

The results indicate that when the location of the pre-existing crack
differs from the primary cracking mode, it tends to accelerate crack
propagation. Conversely, when the pre-existing crack is located in
the same region as the primary cracking mode, its effect on crack
propagation is less pronounced than when it is in the opposite region.
This phenomenon can be attributed to the simultaneous expansion of
pre-existing notches and cracks induced by chemical stress or particle
swelling when located in different regions, thereby increasing the
cracking speed. However, when the cracks are co-located, their effects
overlap, mitigating the severity of crack propagation.

\textbf{Different notch lengths and orientations}

\begin{figure}[!ht]
\centering{}\includegraphics[width=13cm]{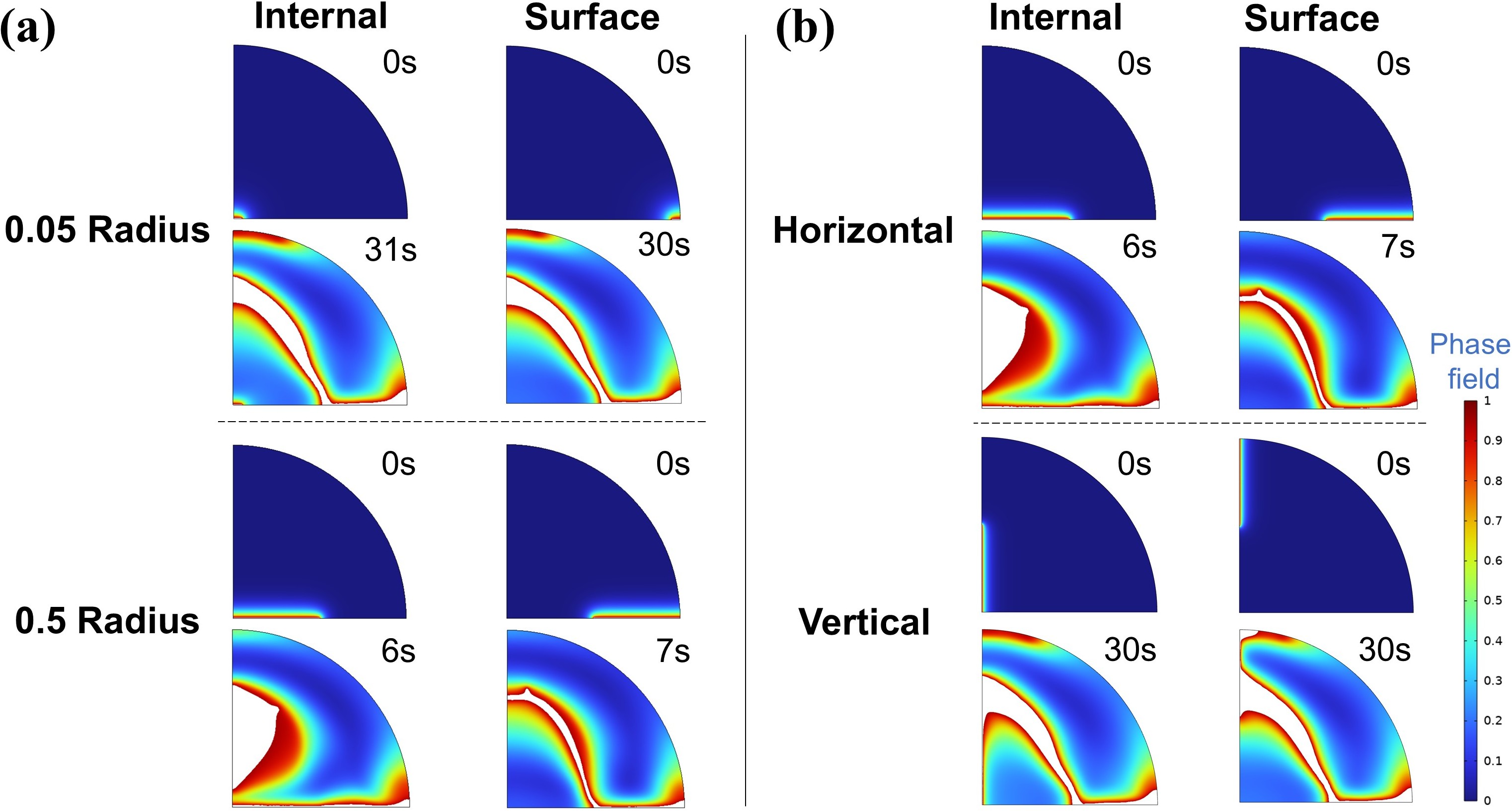}
\centering{}\caption{Fracture time of particles with (a) different pre-existed crack lengths and (b) different pre-existed crack orientations. The particle diameter is 620 nm and the charge rate is 60C.}
\label{fig:9} 
\end{figure}

Subsequently, the effects
of pre-existing crack lengths and orientations were analysed. As 
can be seen in Fig. \ref{fig:9}(a), an increase in the length of
pre-existing notches from 0.05 to 0.5 of the particle radius resulted
in a decrease in particle fracturing time, with a more pronounced
effect observed in particles containing surface cracks. In terms of
crack orientation, as shown in Fig. \ref{fig:9}(b), vertical
pre-existing notches led to a longer particle fracturing time. This
phenomenon can be attributed to the symmetry plane and rotational
axis settings. Horizontal cracks form thin planes that readily expand
at sharp inner boundaries, while vertical cracks form needle-like
structures that are less likely to expand at these boundaries.

\subsection{Fatigue damage}

To study the fatigue damage, cyclic simulations with small charge rates (0.5C - 3C) were performed and a small particle diameter, 200 nm, was selected, which is the usual size of commercial Si nanoparticles. Here, the particle was assumed to be pre-lithiated amorphous Si, which is a common pretreatment in practice \cite{zhang2021challenges}. Then, the effects of pre-existing notch length and charge rate were studied.

\textbf{Effects of pre-existing notch length} 

\begin{figure}[!ht]
\centering{}\includegraphics[width=13cm]{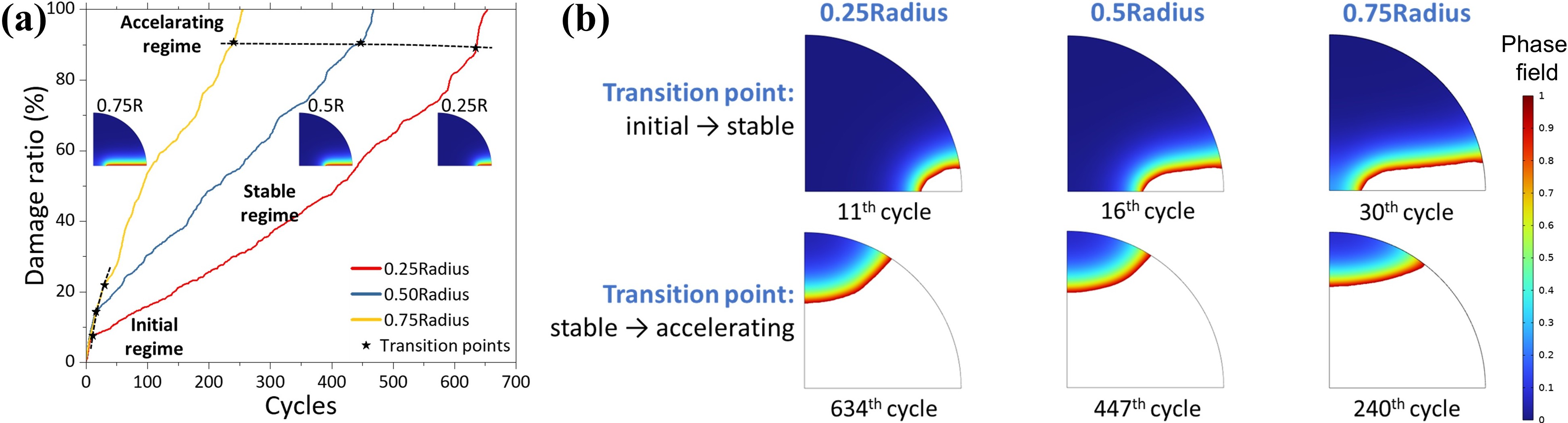}
\centering{}\caption{(a) damage ratio evolution with cycles and (b) fracture behaviours at regime transition points of particles with varying pre-existing notch lengths. The particle diameter is 200 nm, the change rate is 0.5C, and the pre-existing notch locates on the particle surface.}
\label{fig:10} 
\end{figure}

Fig. \ref{fig:10}(a) illustrates the evolution of damage ratios as a function of cycle number for particles with varying pre-existing notch lengths. The damage progression can be categorised into three distinct regimes based on damage rate:  (I) initial regime, which is characterised by a steep increase in the damage ratio during early cycles, driven by stress redistribution as mechanical and chemical (Li-ion concentration) gradients equilibrate. (II) stable regime which exhibits a comparatively low damage rate due to stabilised stress distribution and gradual Li-ion diffusion.  (III) accelerating regime, which is marked by a rapid, nonlinear increase in damage once the critical damage ratio (\~90\% for particles cycled at 0.5C) is exceeded, culminating in catastrophic failure.  The fracture behaviours at regime transition points of particles with varying pre-existing notch lengths were plotted in Fig. \ref{fig:10}(b). As notch length increases, the transition from the initial to stable regime shifts to later cycles and higher damage ratios. This occurs because longer notches more effectively hinder spatially homogeneous Li-ion diffusion, destabilising stress redistribution and thus prolonging the initial regime \cite{pluvinage2003notch}. Also, this diffusion hindrance induces heterogeneous ion concentrations that amplify stress localisation at crack tips. Elevated local stresses accelerate deformation and damage accumulation from the first cycle, thereby delaying the transition to higher damage ratios. 

\textbf{Effects of charge rate}

\begin{figure}[!ht]
\centering{}\includegraphics[width=13cm]{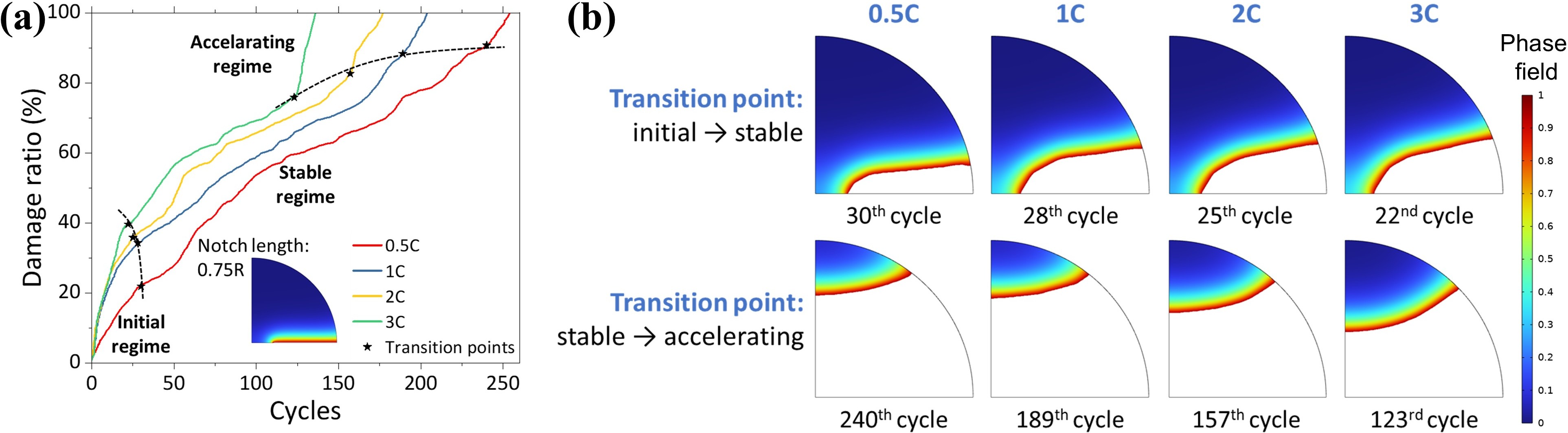}
\centering{}\caption{(a) damage ratio evolution with cycles and (b) fracture behaviours at regime transition points of particles under varying charging rates. The particle diameter is 200 nm and the pre-existing surface notch length is 0.75 radius.}
\label{fig:11} 
\end{figure}

Fig. \ref{fig:11}(a) and (b) respectively show the evolution of damage ratios with cycle number and fracture behaviours at regime transition points of particles under varying charging rates. With increasing charge rate, the transition from the initial regime to the stable regime shifts to earlier cycles and higher damage ratios. This occurs because faster charging intensifies Li-ion flux gradients, amplifying stress concentrations per cycle and hastening the regime transition. Furthermore, the pre-existing notch acts as a stress concentrator, and elevated charge rates exacerbate localised Li depletion/accumulation near the notch. This promotes damage accumulation during early cycles, thus causing a higher damage ratio at the transition points. The diminished critical damage ratio to accelerating regime under faster charging is attributed to severe stress localisation near the pre-existing notch, which elevates tensile strain energy density \citep{gao2021stress}. Consequently, the system reaches the fracture energy barrier at a lower global damage ratio. 
\subsection{Effects of nanopores on mitigating fracture}

\begin{figure}[!ht]
\centering{}\includegraphics[width=13cm]{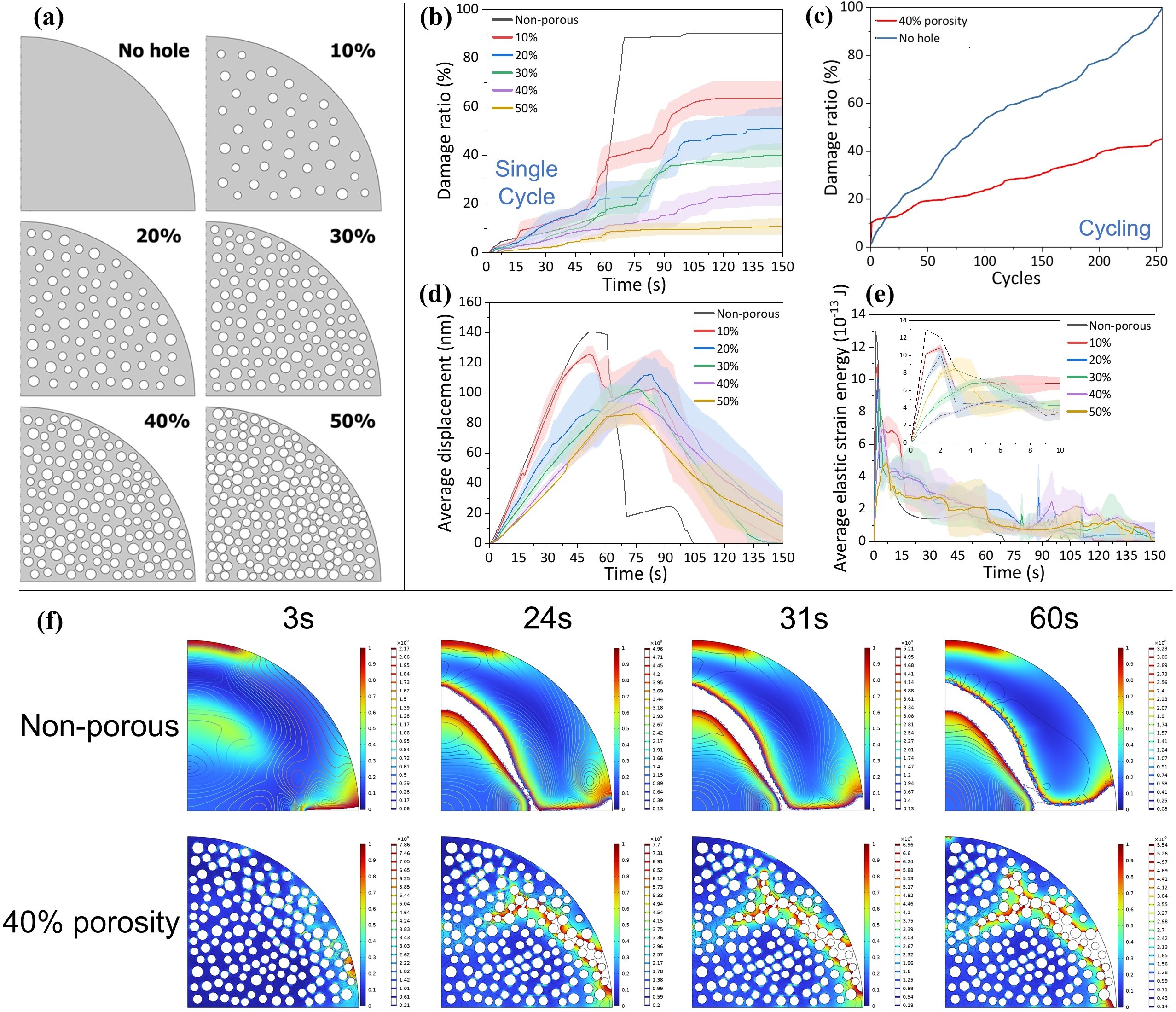} 
\centering{}\caption{(a) geometric model of porous Si particle with different porosities. (b) Normalised crack volume evolution of different porosities in single cycle study, (c) Damage ratio evolution of porous Si particle with a 40\% porosity and non-porous particle in cycling study. (d) Average diameter change and (e) average elastic strain energy of different porosities. (f) Fracture behaviours with stress contour of non-porous particle and porous particle with a 40\% porosity. In the single cycle study, the particle diameter is 620 nm and the charge rate is 60C. In the cycling study, the particle diameter is 200 nm, the charge rate is 0.5C and the pre-existed surface notch length is 0.75Radius. }
	\label{fig:12} 
\end{figure}

Reducing particle expansion is critical for mitigating particle fracturing.
As a solution, porous Si particles were employed in this study. Python
scripts were developed and executed in the application mode of COMSOL Multiphysics to generate randomly and uniformly distributed pores within the particles (Fig. \ref{fig:12}(a)). The internal surfaces of all nanopores are assigned traction-free and no-flux boundary conditions, ensuring that no mechanical load is transferred and no lithium ions pass through the nanopores. 

The cracking curves (Fig. \ref{fig:12}(b)) demonstrated that with
increasing porosity,  the final damage ratio
decreased. As for the cracking speed, for particles with 10\% and 20\% porosity, however, the particles exhibited
faster cracking within the first 60 seconds, which was followed by
a reduction in cracking speed. This rapid initial cracking was likely
caused by the relatively large expansion and the pores, which facilitated
pores connection with pre-existing notches, accelerating early-stage propagation. The porous silicon particle with a 40\% porosity was applied as an example in the cyclic fatigue study. As shown in Fig. \ref{fig:12}(c), the porous particle demonstrated a 52\% reduction in damage accumulation compared to non-porous counterparts, exhibiting a damage ratio of 48\% at the point of complete fracture (100\% damage) in the non-porous particle. The enhanced fatigue resistance of the porous Si particle was also observed in Wei Wang, et al.'s study \citep{wang2015monodisperse}.

To better understand the mechanism by which pores mitigate particle
fracturing, the average displacement and maximum stress of particles
with varying porosities in the single cycle study were analysed, as shown in Fig. \ref{fig:12}(d)
and (e), respectively. The results show that as porosity increases,
the average displacement, the displacement rate decreases, and the average elastic strain energy reduces. These findings suggest that pores within particles can mitigate expansion and dissipate strain energy. Fig. \ref{fig:12}(f) compares the fracture behaviours and stress contours of non-porous particles with those of particles exhibiting 40\% porosity. The analysis reveals that stress concentrations occur predominantly at pore edges, while a lower stress gradient is observed near pores compared to that observed for non-porous particles. Furthermore, the crack propagation path in porous particles is notably redirected and elongated relative to non-porous counterparts. These phenomena indicate that nanopores act as localised stress concentrators, effectively dissipating global tensile stresses and diverting crack propagation through voids to elongate the crack path, thereby slowing crack velocity and prolonging and even preventing fracture. Notably, after completing a full loading cycle (150s), the porous particle remained unfractured, whereas the non-porous particle had fractured at 31s. 

For the economic feasibility of porous silicon particle, while porous silicon anodes cost more than graphite, their high energy density and emerging scalable production methods, e.g., magnesiothermic reduction of silica \citep{entwistle2018review}, justify the investment. A trade-off exists regarding particle size: 620 nm particles exhibit a higher damage ratio than <150 nm nanoparticles but are far cheaper to produce. Specifically, the complex synthesis methods \citep{amoruso2004generation, saito2018solution,bishoyi2022synthesis} of 150 nm particles are much more expensive than the milling or reduction techniques \citep{entwistle2018review,yan2022scalable,kuntyi2022porous} used for 620 nm particles.

\section{Conclusions}

\label{sec5} 
In this study, a multiphysics model fully coupling mass transport, deformation, phase field, and fatigue damage was developed to investigate the cracking and fracturing behaviours of Si particles during single lithiation/delithiation cycles and fatigue damage during multiple cycles. The effects of particle diameter, charge rate, and pre-existing notches were systematically analysed, and a counter map of cracking and fracturing behaviours was built up. Additionally, the influence of pre-existing notch length and charge rate on fatigue damage was examined. Finally, nanopores were introduced into the Si particles as a strategy to mitigate fracturing, and the underlying alleviation mechanism was elucidated, along with the effects of porosity. The key findings are summarised as follows:

1. Cracks in the Si particles initiated due to Li-ion concentration disparity and propagated due to high local stresses at the crack tip.

2. Increasing particle diameter and charge rate altered the crack initiation mode, accelerated crack initiation and particle fracture, and resulted in more severe surface cracks compared to internal cracks.

3. In the single-cycle study, four distinct cracking behaviours were identified: no crack, internal crack, surface crack, and fracture, and a contour map of these behaviours was developed. When the particle diameter is smaller than 150 nm, there is no crack.

4. Pre-existing notches accelerated particle fracture, particularly longer and laterally oriented cracks. When located on the opposite side of crack initiation, pre-existing notches significantly accelerated fracturing, whereas those on the same side had a small impact.

5. In the fatigue damage, three damage regimes were identified: initial regime, stable regime and accelerating regime. Longer pre-existing notch length and larger charge rate shorten the particle cyclic life as a result of a larger localised stress on the boundary, especially the tip, of the pre-existing notch.

6. Nanopores in Si particles mitigated fracture by reducing expansion, dissipating global tensile stresses and elongating the crack propagation path. 

The developed computational framework established a predictive relationship between stress-diffusion coupling theory and particle-level degradation, guiding future design and manufacturing of failure-resistant Si-based anodes for lithium-ion batteries. 

\section{Acknowledgement}
Jie Yang acknowledges the financial support from QMUL/CSC scholarship.

\bibliography{Reference}

\end{document}